\documentclass[conference]{IEEEtran}
\makeatletter
\def\ps@headings{%
\def\@oddhead{\mbox{}\scriptsize\rightmark \hfil \thepage}%
\def\@evenhead{\scriptsize\thepage \hfil \leftmark\mbox{}}%
\def\@oddfoot{}%
\def\@evenfoot{}}
\makeatother
\newtheorem{thm}{Theorem}
\newtheorem{lemma}{Lemma}
\newtheorem{proposition}{Proposition}

\newtheorem{asmption}{Assumption}

\def\isp{\text{ISP}}
\def\cp{\text{CP}}

\usepackage{epsf}
\usepackage{amssymb}
\usepackage{graphicx}
\usepackage{amsmath}
\usepackage[english]{babel}
\usepackage{mathrsfs}
\usepackage{subfigure}

\begin{document}

\title{Network Non-Neutrality through
Preferential Signaling}

\author{Manjesh Kumar Hanawal$^{\dagger \ast}$ and Eitan Altman$^{\dagger}$
\\

{$\dagger$ INRIA B.P.93, 2004 Route des Lucioles
06902 Sophia-Antipolis Cedex, France }\\
{$\ast$ LIA, University of Avignon, 339, chemin des Meinajaries, Avignon, France}\\
Emails: \{mhanawal,eitan.altman\}@sophia.inria.fr }

\maketitle

\begin{abstract}
One of the central issues in the debate on  network neutrality
has been whether one should allow or prevent preferential treatment by
an internet service provider (ISP) of traffic according to its origin. This raised the question of whether to allow an ISP to have exclusive agreement with
a content provider (CP).
In this paper we consider discrimination in the opposite direction.
We study the impact that a CP can have on the benefits of several competing ISPs by sharing private information  concerning the demand for its content.
More precisely, we consider ISPs that compete over access to one
common CP.  Each ISP selects the price that it charges its
subscribers for accessing the content.
The CP is assumed to have private information about demand for its
content, and in particular, about the inverse demand function
corresponding to the content. The competing ISPs are
assumed to have knowledge on only
the statistical distribution of these functions. We derive
in this paper models for studying the impact that the CP can have
on the utilities of the ISPs by favoring one of them by
exclusively revealing its private information. We also consider the case where CP can charge ISPs for
providing such information. We propose two mechanisms based on
{\em weighted proportional fairness}
for payment between ISPs and CP. Finally, we compare the social utility resulting
from these mechanisms with the optimal social utility by introducing a performance metric termed as {\em price of partial bargaining}.
\end{abstract}

\section{Introduction}

The past few years have seen much public debate and legislation initiatives concerning access to the global Internet. Some of the central issues concerned the possibility of discrimination
of packets by service  providers according to their source or destination, or the protocol used.  A discrimination of a packet can occur when preferential  treatment is offered to it either
in terms of the quality of service it receives, or in terms of the cost to transfer it. Much of this debate took part in anticipation of the legislation over ``Net Neutrality", and several public consultations were launched in 2010 (e.g. in the USA, in France and in the E.U.). Network neutrality asserts that packets should not be discriminated. Two of the important issues concerning discrimination of traffic are whether (i) an internet service provider (ISP) may or may not request payment from a content provider (CP) in order to allow it reach its end users, and (ii) whether or not an ISP can have an exclusive agreement with a given CP resulting in a vertical monopoly. Indeed, for Hahn and Wallsten \cite{Economistsvoice06_TheEconomicsOfNet_HahnWallsten}, net neutrality ``usually means that broadband service providers charge consumers only once for internet access, do not favor one content provider over another, and do not charge content providers for sending information over broadband lines to end users".

The network neutrality legislation will determine much of the socio-economic role of the Internet in the future. The Internet has already had a huge impact not only on economy, but also on the exercise of socio-cultural freedom. Directive 2002/22/EC of the European Union, as amended by the Directive 2009/136/EC, established internet access as a universal service\footnote{A universal service has been defined by the EU, as a service guaranteed by the government to all end users, regardless of their geographical location, at reasonable quality and reliability, and at affordable prices that do not depend on the location.}. However, internet is a conglomeration  of several profit making entities. Interaction among these entities is largely governed by economic interests, and their decisions can adversely impact the socio-economic role of the Internet. Thus, it is necessary to understand the interplay between various agents involved, and the knowledge gained can be used in enabling laws that benefits society and its economic development.

This paper pursues a line of research that we have been carrying on for modeling exclusive agreements  between service and content providers and study their economic impact. Such agreements are often called ``vertical monopolies". In some branches of industry, steps have been taken against vertical monopolies. As a result, several railway companies in Europe had to split the railway infrastructure activity from the transportation activity. However, in the telecommunication industry impact of vertical monopolies is not yet clear. The international community is still debating the laws to regulate, or not to regulate, interaction between various agents in the Internet.

In this paper we study another form of nonneutrality\footnote{The traditional net-neutrality discussion is about $\isp$s discriminating $\cp$s by giving them preferential treatment.} resulting from vertical monopolies that arises when a $\cp$ provides private information to an $\isp$. The private information could be popularity of its content, profiles of users interested in different types of content, traffic characterization, usage pattern, etc. We assume that the $\cp$'s private information is related to the demand generated through the $\isp$s.
If CPs can share this private information with an ISP, then that ISP can adopt a more efficient pricing policy than its competitors. For example, recent acquisition of Dailymotion by France T\'el\'ecom (an ISP) enables it to have exclusive information about demand for its video content. We derive game theoretic models that enable to compute the impact of such discrimination on the utility of the ISPs. We model the interaction between the $\isp$s and a $\cp$ as a game, where the CP can share its private information through signals.  We also look at the possibility of $\isp$s paying $\cp$ for access to its private information and study mechanism to decide these payments.

\noindent
\textbf{Related Work:} We have used in the past game theoretical models to study two aspects of vertical monopolies. In \cite{NETCOOP10_NonneutralNetwork_AltmanHanawal} and \cite{WPIN12_GameTheoreticAnalysis_HanawalAltmanSundaresan},
we studied the impact of collusion between an ISP and a CP by jointly determining the price each one charges. We evaluated the impact of such collusion, both on
the colliding companies as well as on the benefits of other ISPs and CPs. In \cite{Netorking12_CompetitionInAccessToContent_JimenezHayelAltman} we studied the impact that an ISP can have by proposing preferential quality of service or cheaper
prices for accessing  a CP with which it has an exclusive agreement. We refer the reader to \cite{GameNets11_NetNeutralityAndQuality_AltmanRojasWong} for a survey on net neutrality debate.  In \cite{GAEB10_SignalingQualityThrough_JanssenRoy},
the authors study a signaling game between high quality and low quality firms in a Bertrand oligopoly \cite{Book_OlogopolyPricising_Vives}.
The quality of each firm is a private information which is signaled to others by the price set on their products.
In \cite{STOC06_TheEffectOfCollusion_HayranpetyanTardosWexler}, the authors propose a metric called {\em price of collusion} to study impact of collusion. In \cite{NETCOOP11_RevisitingCollusionInRouting_AltmanKamedaHayel}, similar definitions are proposed to consider several other scenarios. The authors in \cite{INFOCOM12_HowGoodIsBargained_BlocqOrda} study cooperation in routing games using Nash bargaining solution concept. They study degradation in network performance by introducing a metric called {\em price of selfishness}. Nash bargaining solution is also used in \cite{NEREC09_BargainingPower_Saavedra} to study contracts in nonneutral networks. \\
\noindent
\textbf{Our contributions:} In this paper, we propose a simple model with one $\cp$ and several $\isp$s to analyze a network with vertical monopolies.
\begin{itemize}
\item We first consider the neutral network where the $\cp$ shares private information with all the $\isp$s or none of them. We compare this case with a nonneutral network where the $\cp$ colludes with one of the $\isp$s, say $\isp_1$, and provides signal only to $\isp_1$. We show that an $\isp$ receiving signal improves its monetary gains, while $\cp$ may not.
\item We then consider a case where the $\cp$  charges the $\isp$s for sharing private information. We show that the colluding pair, i.e., the $\cp$ and $\isp_1$ that obtains signal on payment, may not always gain. We characterize the price, that colluding $\isp$ pays to the $\cp$, that results in collusion beneficial to the colluding pair.
\item We then propose mechanisms based on weighted proportional fairness criteria for deciding payments that the colluding $\isp$ makes to the $\cp$ to obtain signal. We compare the social utilities induced by these mechanisms with the optimal social utility by introducing a metric termed as {\em price of partial bargaining}.
\end{itemize}

The paper is organized as follows: In Section \ref{sec:model}, we introduce the model and set up the notations. In Section \ref{sec:NeutralBehavior}, we consider the neutral network in which the $\cp$ provides signals either to all the $\isp$s or to none of them. In section \ref{sec:SignalLinearDemand}, we study the competition assuming the demand generated through ISPs is linear in the user price. Section \ref{sec:nonneutralbehavior} studies a nonneutral behavior in which the $\cp$ colludes with one of the $\isp$s. In Section \ref{sec:signalforprice}, we allow the $\cp$ to charge the colluding $\isp$ for providing the signals. In Section \ref{sec:mechanismsforpricing}, we consider two mechanisms to determine the payment between the colluding pair. In Section \ref{sec:PriceOfPartialBargaining}, we propose a new metric to compare social utilities induced by these mechanism. Finally, we end with conclusions in Section \ref{sec:Conclusion}. All the proofs appear in appendix.

\section{Model}
\label{sec:model}
\begin{figure}[t]
\centering
\includegraphics[width=1.9in, height=1.8in]{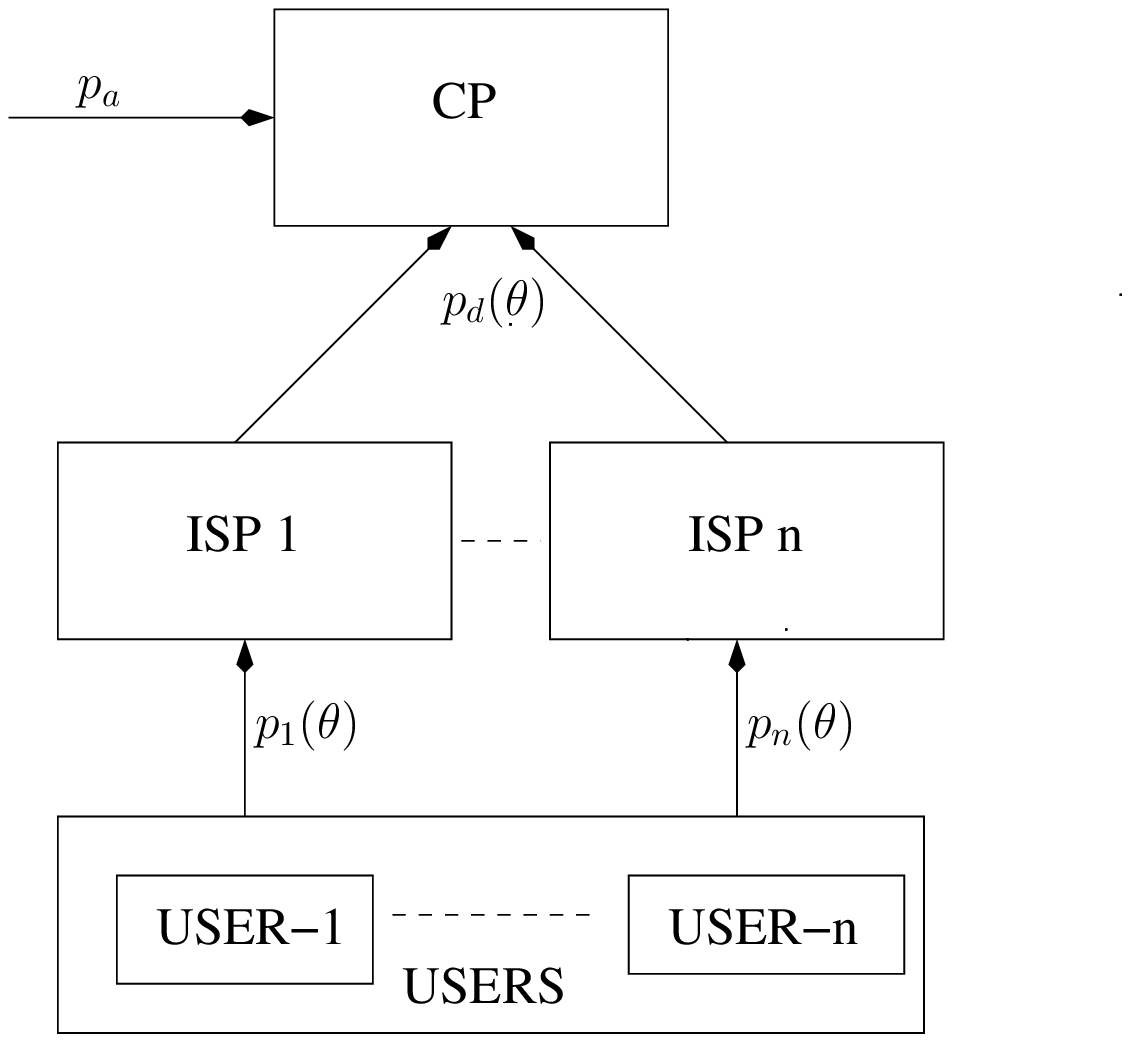}
\caption{Monetary flow in a nonneutral network.}
\label{fig:cp-isp}
\end{figure}

Consider $n$ competing internet service providers ($\isp$s),
namely $\isp_i, \;\; i=1,2\cdots, n$, that provide access to a common content provider ($\cp$).
Each $\isp$ determines the price (per unit of content) that it charges its subscribers. In our model we consider  single $\cp$ as few players, like YouTube, Netflix, account for a significant amount of traffic generated in the Internet.
The demand generated by the subscribers of $\isp_i$ depends on the price of all $\isp$s
as well as on some parameter $\theta$ reflecting private information of the $\cp$. We assume that $\theta$
takes values in some discrete space $\Theta$. The model is show in
Figure \ref{fig:cp-isp}. We summarize the parameters of the model in Table
\ref{tab:Parameters}.

\begin{center}
\begin{tabular}{c|p{2.6in}}
\hline \hline
\label{tab:Parameters}
Parameter & Description \\ \hline \hline
$\theta$             &  indicator of private information of the CP (signal). \\
$p_i(\theta)$        & Price per unit demand charged by $\isp_i$ to its users; this can be a function of  $\theta$. \\
$d_i(\cdot)$         & Demand generated by $\isp_i$. It is a function of the price set by all the $\isp$s and $\theta$. \\
$p_a$                & Advertising revenue per unit demand, earned by the CP. This satisfies $p_a \geq 0$. \\
$p_d$         & Price per unit demand paid by the $\isp$s to the $\cp$ for providing signals. \\
$U_{\isp_i}$         & The revenue or utility of the $\isp_i$.  \\
$U_{\cp}$            & The revenue or utility of the $\cp$. \\
$\gamma$             & Bargaining power of the $\isp$s with respect to the $\cp$. This satisfies $0 < \gamma < 1$. \\
$n$          & Number of $\isp$s. \\
\hline \hline
\end{tabular}
\end{center}
Let the vector $\mathbf{p}(\theta):=(p_1(\theta), p_2(\theta),\cdots,p_n(\theta))$ denote the price set by all the $\isp$s when the signal is $\theta$. We write the demand generated by the subscribers of $\isp_i$ as
\[
  d_i(\theta, \mathbf{p}(\theta))=d_i ( \theta , p_i(\theta) , \mathbf{p}_{-i}(\theta) ) \;\; i=1,2\cdots, n \;\;\text{where}
\]
\[
\mathbf{p}_{-i}(\theta)=(p_1(\theta),p_2(\theta),\cdots,p_{i-1}(\theta),p_{i+1}(\theta), \cdots,p_n(\theta)).
\]
We shall assume that for each $\theta$ and $i$ the demand functions are twice differentiable and satisfy the following monotonicity properties
\begin{equation}
\label{eqn:SignalModel_MonotonicityAssumption}
\frac{\partial d_i(\theta, \mathbf{p}(\theta))}{\partial p_i}<0 \;\text{and}\;\frac{\partial d_i(\theta,\mathbf{p}(\theta))}{\partial p_j}\; > 0\; \text{for} \;j\neq i.
\end{equation}
These conditions imply that if $\isp_i$ increases the access price $p_i$ then subscribers of $\isp_i$ can shift to other $\isp$s, decreasing the demand generated through $\isp_i$ while increasing that generated from the other $\isp$s. Above conditions are common in modeling demand functions in a price competition \cite{Informs04_AGeneralEquilibriumModel_BernsteinFedergruen}. The $\cp$ is assumed to have knowledge on the exact value of $\theta$. The probability $P(\theta)$ that the  private information is of type $\theta$ is a common knowledge to all $\isp$s.

The utility of $\isp_i$ is assumed to have the form
\[
U_{\isp_i} ( \theta , p_i(\theta) , \mathbf{p}_{-i}(\theta) ) = p_i(\theta) d_i ( \theta , p_i(\theta) ,  \mathbf{p}_{-i}(\theta) ).
\]
The $\cp$ earns a fixed advertisement revenue of $p_a$ per unit demand. The total revenue earned by the $\cp$ depends on the effective demand generated by all the $\isp$s. Utility of the $\cp$ is given by
\[
U_{\cp}(\theta, \mathbf{p}(\theta))=\sum_{i=1}^n d_i( \theta , \mathbf{p}(\theta) )p_a.
\]
If $\isp_i$ does not know the actual signal $\theta$ then it can set the price knowing only the distribution of $\theta$. In this case we denote the price by simply $p_i$ (doesn't depend on particular realization of $\Theta$). With some abuse of notation, we denote the utilities of $\isp_i$ as $U_{\isp_i}(\cdot)$ in both the cases. It should be clear from the context if an $\isp$ obtains signals or not. In the current setting $\cp$ acts as a passive player. It can only provide signals to the $\isp$s, but does not control any prices. Its revenue is influenced by the prices set by the $\isp$s. Again, with some abuse of notation we denote the utility of $\cp$ as $U_{\cp}$ in all the cases. 

The demand functions defined above are quiet general. To study the price competition between the $\isp$s we further assume that the demand function $d_i(\theta, \mathbf{p}(\theta))$ is supermodular for each $i$ and $\theta$, and satisfy `dominant diagonal' property. For a twice differentiable function supermodularity property is equivalent to the condition
\noindent
\begin{asmption}[Supermodularity, \cite{Econometrica90_RationalizabilityLearningAnd_MilgromRoberts}]
 \label{asm:SignalSupermodularity}
 \[\frac{\partial^2 d_i(\theta, \mathbf{p})}{\partial p_i \partial p_j} \geq 0 \;\;j\neq i.\]
\end{asmption}
\label{asm:SignalDominantDiagonal}
The dominant diagonal property is defined for all $\gamma, \theta \in \Theta$ as
\begin{asmption}[dominant diagonal]
\[\sum_{j=1}^n d_i(\theta, \mathbf{p})\frac{\partial^2 d_i(\gamma, \mathbf{p})}{\partial p_i \partial p_j}-\frac{\partial d_i(\theta, \mathbf{p})}{\partial p_i}\frac{\partial d_i(\gamma, \mathbf{p})}{\partial p_j} \leq 0 .\]
\end{asmption}
For simplicity, we also assume that the price charged by $\isp_i$ is bounded, say by $p_i^{\max}$ for all $\theta$, such that demand from for all the $\isp$s is positive. Also, the price sensitivity of the subscribers is the same for all the $\isp$s. If $\isp_i$ increases access charges while the others maintain their price, then a fraction of the subscribers move from $\isp_i$ to other $\isp$s without assigning preference to any particular $\isp$. Thus demand function of all the $\isp$s is symmetric.

In the next two sections we study price competition in neutral and nonneutral networks. We define utility and objectives of all the players and compare their revenues in each cases.

\section{Neutral behavior}
In this section we study price competition in a neutral network. In the neutral regime the $\cp$ does not discriminate between the $\isp$s: It shares private information about its content with all the $\isp$s or  none of them. We study these two cases separately, and analyze the impact of having the information on the expected utility of each $\isp$ and the $\cp$ at equilibrium.
\label{sec:NeutralBehavior}
\subsection{No information}
\label{subsec:NeutralNoInfo}
We first consider the neutral behavior in which no information
is shared with the ISPs. The ISPs set their prices knowing only distribution $P(\theta)$.  Recall that in this case we denoted the price charged by $\isp_i$ as
$p_i$. The objective of $\isp_i$ is to set $p_i$ that maximizes its expected utility, i.e.,
\begin{eqnarray*}
E[U_{\isp_i}]:=E_{\theta}[U_{\isp_i}(\theta, p_i,\mathbf{p}_{-i})]= \int P ( d \theta ) p_i d_i ( \theta , p_i , \mathbf{p}_{-i} ).
\end{eqnarray*}
where the operator $E_{\theta}[\cdot]$ denotes expectation with respect to the random signal $\theta$.

\subsection{Full information}
\label{subsec:NeutralFullInfo}
Let us consider the case where the $\cp$ gives signals to all the $\isp$s, i.e., all the $\isp$s are given $\theta$. We also assume that the signal is sent to all the $\isp$s simultaneously. Note that the $\cp$ providing signals to all the $\isp$s is a non-discriminatory act. Hence we consider this case under neutral regime.

$\isp$s can use knowledge of $\theta$ to set the price charged form their users.
The objective of $\isp_i$ is to maximize its expected utility given by
\begin{eqnarray*}
E[U_{\isp_i}] &=& E[U_{\isp_i}(\theta, p_i(\theta),\mathbf{p}_{-i}(\theta))]\\
&=&\int P ( d \theta )U_i(\theta, p_i(\theta),\mathbf{p}_{-i}(\theta)).
\end{eqnarray*}
Note that if any vector $(p_i(\theta))_{\theta \in \Theta}$ maximizes the expected utility for a given $\{p_{-i}(\theta):\theta \in \Theta\}$, then $p_i(\theta)$ also maximizes $U_{\isp_i}(\theta,\mathbf{p}(\theta))$ for each $\theta$. Thus the objective of each player is to maximize $U_{\isp_i}(\theta, p_i(\theta),\mathbf{p}_{-i}(\theta))$ for each value of $\theta$. In this case strategy of each $\isp_i$ is to choose a pricing function on $\Theta$, i.e., $p_i: \Theta \rightarrow [0\; p_i^{\max}]$.

\begin{thm}
\label{thm:SignalNeutralUtiltiyCompare}
Assume that the demand function are supermodular and satisfy the dominant diagonal property for all the $\isp$s and $\theta$. Then the price competition in the neutral regime has the following properties:
\begin{itemize}
\item When all the $\isp$s obtain the signals, equilibrium exists and unique.
\item When none of the $\isp$s obtain the signals, equilibrium exists and unique. $\hfill \IEEEQEDopen$
\end{itemize}
\end{thm}
The proof follows by verifying that the expected utilities are supermodular and satisfy dominant diagonal condition.

\section{Nonneutral behavior}
\label{sec:nonneutralbehavior}
In a nonneutral network the $\cp$ can discriminate between the $\isp$s by giving preferential treatment to, or making an exclusive agreement with, one of the $\isp$s. In this subsection we assume that $\cp$ shares information with one of the $\isp$s. Without loss of generality we assume that the $\cp$ shares information with $\isp_1$ through signals. Then $\isp_1$ can set access price knowing the signal, whereas  the other $\isp$s does so knowing only the distribution. In this case we say that $\cp$ and $\isp_1$ are in collusion, and refer to them as colluding pair. The utilities of $\isp_1$ and other $\isp$s are given, respectively, as follows:
\begin{eqnarray*}
U_{\isp_1}(\theta,p_1(\theta),p_{-1}) &=& d_i(\theta, p_1(\theta),p_{-1}) p_1(\theta)\\
U_{\isp_j}(\theta,p_j,p_{-j})&=&d_j(\theta, p_j, p_{-j})p_j\; j=2,\cdots,n.
\end{eqnarray*}
In the above utilities we write $p_1(\cdot)$ as a function of the $\theta$, whereas $p_j, \; j=2,3,\cdots,n$ are constants chosen knowing only distribution of $\theta$. The objective of $\isp_1$ and $\isp_j, \; j=2,3,\cdots,n$ is to maximize their expected utilities given, respectively, as follow
\begin{eqnarray*}
E[U_{\isp_1}]&=&\int P(d \theta)U_1(\theta,p_1(\theta),p_{-1}),\\
E[U_{\isp_j}]&=&\int P(d \theta)U_j(\theta,p_j,p_{-j}).\\
\end{eqnarray*}

\begin{thm}
\label{thm:SignalNonneutralUtilityCompare}
Assume that the demand functions are supermodular and satisfy diagonal dominance property for each $\isp$ and $\theta$. Assume that the $\cp$ provides information only to $\isp_1$. Then equilibrium exists in the nonneutral regime and is unique. $\hfill \IEEEQEDopen$
\end{thm}

After establishing existence  of equilibrium prices in both neutral and nonneutral regimes it is interesting to compare the utilities in both the regimes to see the impact of sharing private information on revenues of the $\isp$s. Though it appears that an $\isp$  receiving signal should obtain higher revenue compared to an $\isp$ without signal, this observation is not true in general: One can construct simple examples where a player with more information can gain or loose even when equilibrium is unique. For example, see \cite{IJGT03_PositiveValueOfInformation_BassanGossner}. Also, the authors in the same paper give conditions on the feasible payoff sets that ensures higher utilities at equilibrium for a more informed player. However, these conditions are not easy to verify in a game with continuous strategy space.

There are several demand functions which are log-supermodular and satisfy monotonicity and diagonal property. For example, Linear, Logit, Constant Elasticity of Substitution (CES), Cobb-Douglas, etc \cite{Informs04_AGeneralEquilibriumModel_BernsteinFedergruen}. By appropriately choosing the term that depends on the signal $\theta$, we can use any of them to model demand functions. In the following we restrict our attention to linear demand function to study the impact of signaling. Linear demand functions are often used to model demand functions in economic literatures \cite{Informs04_AGeneralEquilibriumModel_BernsteinFedergruen} due to simplicity of analysis.

\section{Linear demand function}
\label{sec:SignalLinearDemand}
For given price vector $\mathbf{p}(\theta)$, assume that demand generated through $\isp_i$ is given by
\begin{equation}
\label{eqn:SignalLinearDemand}
d_i(\theta, \mathbf{p}(\theta))=D_i(\theta)-\alpha p_i(\theta)+\beta \sum_{j\neq i}p_j.
\end{equation}
where $\alpha > 0$ and $\beta >0$, and $D_i(\theta)$ denotes the demand generated through $\isp_i$
when private information of the $\cp$ corresponds to $\theta$ and all the $\isp$s give free access to the $\cp$ . For simplicity, we assume that all the $\isp$s are equally competitive and set $D_i(\theta)=D(\theta)$, i.e., if all the $\isp$s offer free subscription then the demand generated through each $\isp$ is the same. Also, the users are assumed to be equally sensitive to the price set by each $\isp$, thus we take $\alpha$ and $\beta$ to be the same in the demand function of each $\isp$. Note that linear demand function is supermodular and satisfies the dominant diagonal property if $\alpha > (n-1)\beta$. We assume this relation holds in the sequel.

Due to simple structure of linear demand function one can compute the equilibrium prices and equilibrium utilities in both the neutral regime and nonneutral regime explicitly and compare.
\begin{thm}
\label{thm:ISPUtilityComparison}
Assume that demand function is (\ref{eqn:SignalLinearDemand}) for each $\isp$ and $\alpha > (n-1)\beta$. Then, in the neutral regime, $\isp$s obtain higher revenue when all of them receive signals, compared to the case where none of them receive signals.
In the nonneutral regime,
\begin{enumerate}
\item the colluding $\isp$ obtains higher revenue at equilibrium than the noncolluding $\isp$s.
\item the revenue of the colluding $\isp$ further improves if the noncolluding $\isp$s also receive signals.
\item the revenue of the noncolluding $\isp$s remains the same as in the neutral regime where none of the $\isp$s receive signals (no information).
\end{enumerate}
Finally, the revenue of the $\cp$ at equilibrium is the same in all the cases.  $\hfill \IEEEQEDopen$
\end{thm}
As we note from the above theorem, if $\isp$s receive signal they only earn higher revenues. However, this increase in $\isp$s revenue is not because of increased demand, but due to the optimal choice of subscription prices. In particular, as shown in the proof, the demand generated through each $\isp$ remains the same at equilibrium, irrespective of whether it receives signal or not. Thus, the $\cp$ does not gain anything by sharing its private information as its revenue depends only on the total demand generated. Hence the $\cp$ has an incentive to charge the $\isp$s for sharing its private information.

In the rest of the paper we restrict our attention to the case with just two $\isp$s for ease of exposition. However, it is not very restrictive as end users often face a duopoly $\isp$ market.

\section{Signaling with Side Payments}
\label{sec:signalforprice}
In this section we assume that $\cp$ charges $\isp$s for providing signaling information. We assume that if an $\isp$ receives signal from the $\cp$ it pays a fixed price of $p_d$ per unit demand generated by its subscribers. We refer to $p_d$ as side payment. We consider the nonneutral regime where $\isp_1$ gets preferential treatment from the $\cp$. Our aim is to study the impact of side payment on the equilibrium utilities of the colluding pair ($\isp_1$ and $\cp$) and on the non-colluding  $\isp_2$. In particular, we will be interested in characterizing the values of side payment that makes the collusion with the $\cp$ beneficial to $\isp_1$ and vice versa.
\subsection{Nonneutral network with pricing}
\label{subsec:NonnuetralNetwrokWithPricing}
We define the utility of the $\isp_1,$ $\isp_2$ and $\cp$ respectively as follows:
\begin{eqnarray*}
U_{\isp_1}(\theta, p_1(\theta), p_2)&=&(D_0(\theta)-\alpha p_1(\theta)+ \beta p_2)(p_1(\theta)-p_d)\\
U_{\isp_2}(\theta, p_1(\theta), p_2)&=&(D_0(\theta)-\alpha p_2 + \beta p_1(\theta))p_2\\
U_{CP}(\theta, p_1(\theta), p_2)&=&(D_0(\theta)-\alpha p_1(\theta)+ \beta p_2)(p_a+p_d) \\
&+&(D_0(\theta)-\alpha p_2+ \beta p_1(\theta))p_a
\end{eqnarray*}
The $\cp$ informs the value of $p_d$ to $\isp_1$ while they enter in to the agreement. Thus $\cp$ acts as a passive player, providing signals, which in turn affects the demand generated by the subscribers of the $\isp$s. We proceed to analyze the game between $\isp_1$ and $\isp_2$. The objective of each $\isp$ is to maximize its expected utility.

\begin{proposition}
\label{prop:EquilibriumUtiltiyPartialPricing}
In the collusion assume that the $\cp$ imposes price on $\isp_1$ for sharing private information rather than giving it for free. Then, equilibrium revenue of $\isp_1$ decreases, whereas that of the noncolluding $\isp$ increases. $\hfill \IEEEQEDopen$
\end{proposition}
Thus pricing in the nonneutral network have a positive externality on the non-colluding $\isp$. This behavior can be explained as follows: When $\isp_1$ is charged, it too charges its subscribers higher to compensate for the extra payment it makes to the $\cp$ (see proof of Prop. \ref{prop:EquilibriumUtiltiyPartialPricing}).
Whereas  $\isp_2$ does not need to increase its access fee, and also, some of the users of $\isp_1$ shift to $\isp_2$. This increases demand generated through $\isp_2$, thus improving the revenue of the non-colluding $\isp$. Unlike for the non-colluding $\isp_2$, revenue of the colluding pair may or may not improve. It depends on the value of $p_d$. The following theorem characterizes its range.

\begin{thm}
\label{thm:SignalWithPricingPriceRegion}
Assume that in a collusion $\cp$ provides signal to $\isp_1$ by charging a price $p_d$ per unit demand. Also assume that the distribution of $\theta$ is such that $Var(D(\theta))\leq E[D(\theta)]$.  Then, \\
1) $\isp_1$ has an incentive to collude with the $\cp$ if and only if
\[p_d \leq \frac{(2\alpha+\beta) E[D(\theta)]}{2\alpha^2-\beta^2}\left \{1-\sqrt{1-\frac{(2\alpha-\beta)^2 Var(D(\theta))}{4\alpha^2 E^2[D(\theta)]}}\right \}.\]
2) The $\cp$ has an incentive to collude if and only if \begin{equation}
\label{eqn:CondnPatialPriceCPBenefit}
p_d \leq \frac{E[D(\theta)](4\alpha^2-\beta^2)}{(2\alpha-\beta)(2\alpha^2-\beta^2)}
-\frac{(2\alpha^2-\beta^2-\alpha\beta)}{2\alpha^2-\beta^2}p_a
\end{equation}
Further, the colluding $\isp$ obtains higher revenue than the non-colluding $\isp$ if and only if
\begin{eqnarray*}
p_d \hspace{-.2cm}&\leq& \hspace{-.3cm}\frac{(2\alpha+\beta) E[D(\theta)]}{2\alpha^2-\beta^2+ \alpha \beta}\\
\hspace{-.2cm}&\times&\hspace{-.3cm} \left \{1-\sqrt{1-\frac{(2\alpha-\beta)^2 Var(D(\theta))(2\alpha^2-\beta^2+\alpha\beta)}{4\alpha^2 E^2[D(\theta)](2\alpha^2-\beta^2-\alpha\beta)}}\right \}.
\end{eqnarray*} $\hfill \IEEEQEDopen$
\end{thm}

\begin{figure}[t]
\centering
\includegraphics[scale=.5]{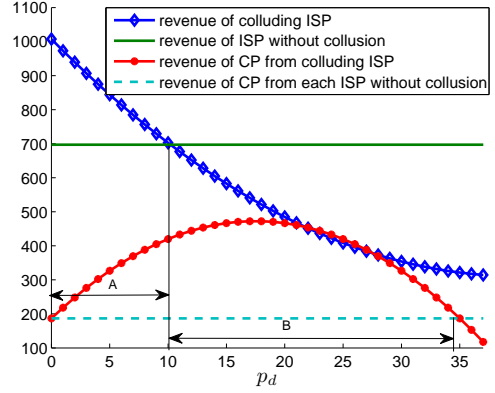}
\caption{Utilities of colluding ISP and CP.}
\label{fig:PriceRegion}
\end{figure}

In Figure \ref{fig:PriceRegion}, we plot the utilities of colluding $\isp$ and $\cp$ as a function of $p_d$. In generating the plot we used the following parameters. The signal $\theta$ takes three values: high (H), medium (M) and low (L), which corresponds to demands $D(H)=200, D(M)=50, D(L)=20$. The distribution of $\Theta$ is taken as $\Pr(\Theta=H)=0.1, \Pr(\Theta=M)=0.6, \Pr(\Theta=L)=0.3$. The other parameters are $\alpha=2$, $\beta=1$ and $p_a=5$. \\

As shown in the figure, if the side payment lies in the
region marked A, then the colluding $\isp$ obtains higher revenue at equilibrium. If it is charged a price outside the region A, then collusion with the $\cp$ is not beneficial to any $\isp$. For the $\cp$, it appears that higher side payment will increase its revenues. But this is not the case. If $\isp$ has to pay higher price to $\cp$ its demand goes down, which in turn reduces the revenue of the $\cp$. Thus it is not beneficial for the  $\cp$ to charge high prices from $\isp$s. Indeed, if $\cp$ charges price beyond region B, governed by (\ref{eqn:CondnPatialPriceCPBenefit}), it will not improve its revenue. The collusion between $\cp$ and $\isp$ is profitable to both if and only if side payment lies in the region A. Also, Note that in the region A, though both $\isp$ and $\cp$ benefit, $\cp$ can obtain higher revenues by increasing the side payment but at the cost of reducing the revenue of $\isp_1$. Thus it becomes important to decide how the side payments should be set so that all the players remain satisfied. In the next section we look for mechanisms to address this issue. \\

%

\section{Mechanisms for setting side payments}
\label{sec:mechanismsforpricing}
In this section we look for mechanisms that take into account the bargaining power (weight) of each player.
As in the previous section, $\isp_1$ is in collusion with $\cp$, in which $\cp$ shares private information with $\isp_1$ on payment. We assume that both $\isp_1$ and $\cp$ decide payment in presence of an arbitrator, and refer to the process as {\em bargaining}. Arbitrator can be a regulating authority, or a disinterested third party who aims to set a side payment that maximizes, in a sense made precise below,  the revenues earned by the colluding pair.
We consider the following two game models.\\
The timing for the first game is as follows.
\begin{itemize}
\item $\isp_1$ and the $\cp$ bargain over the payment $p_d$.
  \item $\isp_1$ and $\isp_2$ set the access price. The prices are set simultaneously .
  \item The subscribers react to the prices and set the demand generated through each $\isp$.
\end{itemize}
In the second game, timing is as follows:
\begin{itemize}
 \item $\isp_1$ and $\isp_2$ set their access price simultaneously.
\item $\isp_1$ and the $\cp$ then bargain over the payment $p_d$.
\item The subscribers react to the prices and set the demand generated through each $\isp$.
\end{itemize}

The first game arises when the private information of the $\cp$ changes over a slower time-scale making the agreement between $\isp_1$-$\cp$ last for longer duration, whereas the $\isp$s vary access fee over a comparatively faster time-scale. The second game arises in cases where private information of the $\cp$ varies over a faster time-scale making the $\isp_1$-$\cp$ to renegotiate the side payments often, whereas the $\isp$s price to its subscribers varies over a slower time-scale. We analyze both models via backward induction and identify the equilibria. In the sequel, we refer to the first game as {\em pre bargaining} game, and the second one as {\em post bargaining} game.

In deciding side payment the arbitrator takes into account only revenue of $\cp$ earned by traffic generated through $\isp_1$. For a given $\{p_1(\theta), \theta \in \Theta\}$ and $p_2$, the arbitrator decides
side payment $p_d$ from $\isp_1$ to the $\cp$ based on weighted proportionally fair allocation criteria given as follows:
\begin{equation}
\label{eqn:NashBargainingOptimalPrice}
  p^*_d \in \arg \max_{p^d} E[U_{\isp_1}]^{\gamma} E[U_{\cp}]^{1 -\gamma}.
\end{equation}
The parameter $\gamma
\in [0,\; 1]$ determines the bargaining power of $\isp_1$ with respect to the $\cp$.

If we take $\gamma = 1/2 $, then the above maximization is equivalent to that of product of the utilities of $\isp_1$ and the $\cp$. This is then the standard proportional fair allocation \cite{ETOT97_ChargingAndRateControl_Kelly} and is based on {\em Nash bargaining solution} which is known to satisfy set of four axioms \cite{Econometrica50_TheBargainingProblem_Nash}. We note the method discussed in this section is a modified version of the standard Nash bargaining solution, abusing terminology we continue to refer to it as bargaining.
\noindent
We may imagine that the bargaining is done by another player, the regulator, whose (log) utility equals
\begin{equation}
  \label{barg}
  \overline{U}_{\textsf{regulator}} := \gamma \overline{U}_{\isp_1} + (1-\gamma) \overline{U}_{\cp},
\end{equation}
where $\overline{U}_{\isp_1} = \log E[U_{\isp_1}]$ and $\overline{U}_{\cp} = \log E[U_{\cp}]$.

Recall that the strategy of $\isp_1$ is to choose price vector $\{p_1(\theta), \theta \in \Theta\}$ and that of $\isp_2$ is to choose $p_2$ knowing only distribution of $\Theta$, and utilities of $\isp_1,$ $\isp_2$ and $\cp$ are defined respectively as follows:
\begin{eqnarray*}
\label{eqn:PricingFullInfoUtilityISP1}
E[U_{\isp_1}]&=&E[(D(\theta)-\alpha p_1(\theta)+ \beta p_2)(p_1(\theta)-p_d)]\\
\label{eqn:PricingFullInfoUtilityISP2}
E[U_{\isp_2}]&=&E[(D(\theta)-\alpha p_2 + \beta p_1(\theta))]p_2 \\
E[U_{\cp}]&=& E[(D(\theta)-\alpha p_1(\theta)+ \beta p_2)](p_a+p_d)   \\
&& + E[(D(\theta)-\alpha p_2 + \beta p_1(\theta))]p_a.
\end{eqnarray*}
\noindent
We now return to our game model where the colluding pair decides side payment in presence of the arbitrator.. In both games the $\cp$ is a passive player. In the first game, $\isp_1$ and the $\cp$ bargain over side
payment and then the $\isp$s set their price competitively. In the second, $\isp_1$ choose price knowing that he will bargain with the $\cp$ subsequently. Our aim is to compare the expected utilities of each player as a function of $\gamma$, and study how the bargaining power influences the players' preference for the bargaining modes, i.e, pre bargaining or post bargaining.

\subsection{Pre bargaining}
\label{subsec:prebargaining}
At the beginning, $\isp_1$ bargain with the $\cp$ and decide side payments. In this bargaining process we take into account the bargaining power of each player. Once the side payment is set, the game is between the two $\isp$s alone who set their prices competitively to maximize their revenue.

\noindent
We computed the equilibrium utilities of  players in the proof of Proposition \ref{prop:EquilibriumUtiltiyPartialPricing} given as follows:
\begin{eqnarray}
\nonumber
\hspace{-.4cm}E [U_{\isp_1}]\hspace{-.4cm} &=& \hspace{-.3cm}\alpha E \left [ \left ( \frac{D(\theta)}{2\alpha}+ \frac{\beta E[D(\theta)]}{2\alpha(2\alpha-\beta)}-\frac{p_d(2\alpha^2-\beta^2)}{4\alpha^2-\beta^2}\right )^2 \right ]\\
\label{eqn:EquilibriumUtiltilityISP1PartialPricing}
\\
\label{eqn:EquilibriumUtiltilityISP2PartialPricing}
E[U_{\isp_2}]\hspace{-.3cm}&=&\hspace{-.3cm}\alpha \left ( \frac{E[D(\theta)]}{(2\alpha-\beta)}+ \frac{\alpha\beta}{4\alpha^2-\beta^2}p_d\right )^2\\
\nonumber
E[U_{\cp}]&=&\frac{2\alpha E[D(\theta)]p_a}{2\alpha-\beta}  -\alpha \frac{(2\alpha^2-\beta^2)}{4\alpha^2-\beta^2}p_d^2 \\
\label{eqn:EquilibriumUtiltiyPartialPricingCP}
&+&  \left ( \frac{\alpha E[D(\theta)]}{2\alpha-\beta}-\frac{\alpha (2\alpha^2-\beta^2-\alpha \beta)p_a}{4\alpha^2-\beta^2}\right )p_d.
\end{eqnarray}
Utility of the $\cp$ given above consists of revenue generated from traffic of both the $\isp$s. The portion that comes from $\isp_1$ is given by
\[\alpha E \left [ \left ( \frac{D(\theta)}{2\alpha}+ \frac{\beta E[D(\theta)]}{2\alpha(2\alpha-\beta)}-\frac{p_d(2\alpha^2-\beta^2)}{4\alpha^2-\beta^2}\right ) \right ]\]
\noindent
Then the optimization problem of the arbitrator is given by
\begin{eqnarray}
\label{eqn:SidePayPreBargainOptimization}
\arg \max_{p_d} \alpha E \left [ \left ( \frac{D(\theta)}{2\alpha}+ \frac{\beta E[D(\theta)]}{2\alpha(2\alpha-\beta)}-\frac{p_d(2\alpha^2-\beta^2)}{4\alpha^2-\beta^2}\right )^2 \right ]^\gamma \hspace{-.35cm}\times
\nonumber\\
E \left [ \left ( \frac{D(\theta)}{2\alpha}+ \frac{\beta E[D(\theta)]}{2\alpha(2\alpha-\beta)}-\frac{p_d(2\alpha^2-\beta^2)}{4\alpha^2-\beta^2}\right ) \right ]^{1-\gamma}\hspace{-.5cm}(p_a+p_d)^{1-\gamma}. \nonumber \\
\end{eqnarray}
Taking logarithm of the objective function and differentiating with respect to $p_d$ it is easy to verify that the above optimization problem has unique solution.

\subsection{Post Bargaining}
\label{subsec:postbargain}
In the second game, $\isp$s set price competitively knowing that the arbitrator will decide side payment between $\isp_1$ and $\cp$ according to (\ref{eqn:NashBargainingOptimalPrice}).

\begin{figure}[t]
\centering
\includegraphics[scale=.6]{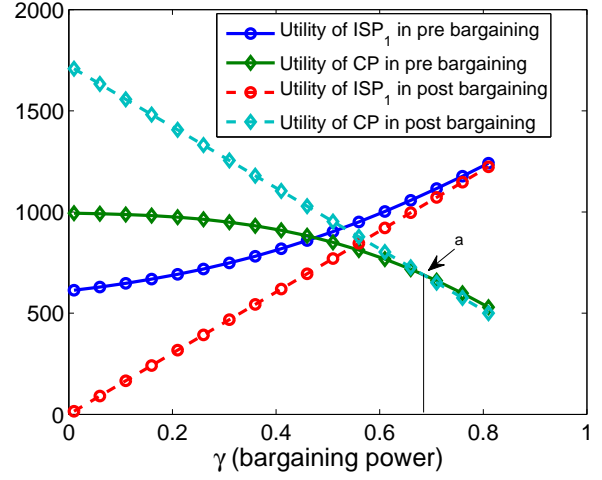}
\caption{$\isp$ and $\cp$ utility in two game models $\alpha=2,\beta=1.5$}
\label{fig:BargainingISPUtilityCOmparison}
\end{figure}

As in the pre bargaining case, we analyze this game in the reverse order, i.e., we first look at the side payment set by the arbitrator as a function of $\isp$ prices, and then study the competition between the $\isp$s.
Note that the demand function $d_2(\theta, p_1(\theta), p_2)$ does not depend on $p_d$. Then the side payment set by the arbitrator for a given $\{p_1(\theta), \theta \in \Theta\}$ and $p_2$ is such that
\begin{eqnarray*}
\label{eqn:BargainingAfterOptimalPrice}
p_d^* \in \arg \max_{p_d} E[d_1(\theta,p_1(\theta),p_2)(p_1(\theta)-p_d))]^{\gamma}\times\\
E[d_1(\theta,p_1(\theta),p_2)(p_a+p_d))]^{1-\gamma}.
\end{eqnarray*}
\begin{lemma}
\label{lma:NeutralPostBargainOptimalSidePayment}
For a given $\{p_1(\theta), \theta \in \Theta\}$ and $p_2$ the arbitrator sets side payment as
\begin{eqnarray*}
p_d^*&=&\frac{E[d_1(\theta,p_1(\theta),p_2)(p_1(\theta))]}{E[d_1(\theta,p_1(\theta),p_2)]}- \frac{E[d_1(\theta,p_1(\theta),p_2)p_a]}{E[d_1(\theta,p_1(\theta),p_2)]}.
\end{eqnarray*}$\hfill \IEEEQEDopen$
\end{lemma}
Substituting this expression of $p_d^*$ in the utility of $\isp_1$ and $\cp$, the modified utilities are as follows:
\begin{eqnarray*}
\overline{U}_{\isp_1}&=&\gamma E[d_1(\theta,p_1(\theta),p_2)(p_1(\theta)+p_a)]\\
\overline{U}_{\cp_1}&=&(1-\gamma) E[d_1(\theta,p_1(\theta),p_2)(p_1(\theta)+p_a)]
\end{eqnarray*}
Note that in the above expressions  $E[d_1(\theta,p_1(\theta),p_2)(p_1(\theta)+p_a)]$ is total revenue earned by both $\isp_1$ and $\cp$ from the traffic generated through $\isp_1$.
In the post bargaining they shares this total revenue in proportion to their bargaining power. We proceed to analyze the game between $\isp$s with the modified utility of $\isp_1$.
\begin{proposition}
\label{prop:NeutralPostBargainingUtilities}
In the post bargaining game with the modified utilities, the equilibrium utilities are as follows:
\begin{eqnarray*}
\nonumber
E[U_{\isp_1}]\hspace{-.3cm}&=&\hspace{-.3cm}\gamma \alpha E \left [\left (\frac{D(\theta)}{2\alpha}+\frac{\beta E[D(\theta)]}{2\alpha(2\alpha-\beta)}+\frac{(2\alpha^2-\beta^2)p_a}{4\alpha^2-\beta^2}\right )^2\right ] \\
E[U_{\isp_2}]\hspace{-.3cm}&=&\hspace{-.3cm} \alpha \left (\frac{E[D(\theta)]}{2\alpha-\beta}-\frac{\alpha \beta p_a}{4\alpha^2-\beta^2} \right )^2 \\
E[U_{\cp}]&=&  \alpha \left (\frac{E[D(\theta)]}{2\alpha-\beta}-\frac{\alpha \beta p_a}{4\alpha^2-\beta^2} \right )p_a + \\
(1-\gamma)\alpha \hspace{-.3cm} &E& \hspace{-.3cm} \left [\left (\frac{D(\theta)}{2\alpha}+\frac{\beta E[D(\theta)]}{2\alpha(2\alpha-\beta)}+\frac{ (2\alpha^2-\beta^2)p_a}{4\alpha^2-\beta^2}\right )^2\right ].
\end{eqnarray*}$\hfill \IEEEQEDopen$
\end{proposition}
We can now compare the expected utilities of the players in the two games obtained in Proposition \ref{prop:NeutralPostBargainingUtilities} and equations (\ref{eqn:EquilibriumUtiltilityISP1PartialPricing})-(\ref{eqn:EquilibriumUtiltiyPartialPricingCP}). First note that the expression for the utility of $\isp_1$ and the $\cp$ in the pre bargaining game is similar to that in the post bargaining game with $p_d$ replaced by $-p_a$. Also, as seen from (\ref{eqn:SidePayPreBargainOptimization}) the side payment set in the pre bargaining game satisfies $p_d > - p_a$. This gives the impression that $\isp_1$ prefers the post bargaining mechanism in setting side payment. However, note the multiplicative factor $\gamma$ in the utility of $\isp_1$ in the post bargaining game. When the bargaining power of $\isp_1$ is small, it gets only a small fraction of the joint revenue it earns with the $\cp$. Thus $\isp_1$ prefers pre bargaining to decide side payment, whereas $\cp_1$ prefers post bargaining. As the bargaining power of the $\isp_1$ increases, $\cp$ gets only a smaller fraction of the total revenue earned, and it prefers post bargaining. We plot the expected utility of $\isp_1$ and $\cp$ in both the game models as a function of $\gamma$ in Figure \ref{fig:BargainingISPUtilityCOmparison}.
In generating the plots we used the same parameters as in Figure \ref{fig:PriceRegion} with $\beta=1.5$. As seen, there is a threshold on the bargaining power, marked as point $a$ in the figure, below which $\cp_1$ prefers pre bargaining and above post bargaining.

\section{Price of Partial Bargaining}
\label{sec:PriceOfPartialBargaining}
In the previous section we studied mechanisms to decide the payment based on weighted proportionally fair criteria. Another natural choice is to set a side payment such that the sum of the utility of all the players is maximized at equilibrium. Let $\tilde p_d$ denote this side payment, i.e.,
\[\tilde p_d \in \arg \max_{p_d}\quad E[U_{\isp_1}]+E[U_{\isp_2}]+E[U_{\cp}],\]
where the utilities are computed at the equilibrium prices of the players. We denote the expected sum of equilibrium utilities calculated at $\tilde p_d$ as $\tilde U$. Recall that we denoted the side payment obtained in weighted proportional fairness solution as $p_d^*$ in (\ref{eqn:NashBargainingOptimalPrice}).
Let the expected sum of the equilibrium utilities calculated at the weighted proportional fairness solution be denoted as $\overline U$. We will be interested in studying how good $\overline U$ is compared to $\tilde U$.

In this section we do not take into account the bargaining power of each player. We shall be interested in simply the product of utilities (i.e., without the exponents in (\ref{eqn:NashBargainingOptimalPrice})). A more interesting analysis would be to compare optimal $\alpha$-fair social equilibrium utility, interpreting fairness factor as the bargaining power. However, we will not pursue this thought in this work.

\begin{figure}[t]
\centering
\includegraphics[scale=.5]{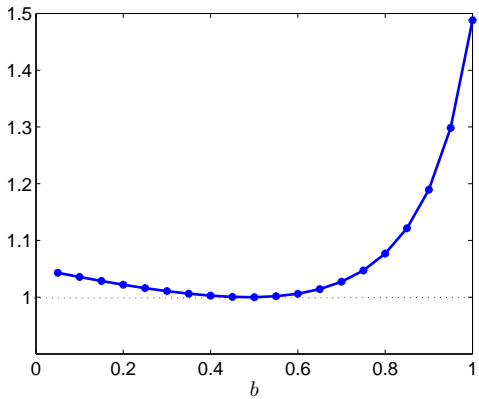}
\caption{Price of partial Bargaining $\alpha=2$}
\label{fig:PoPBPreBargain}
\end{figure}

In \cite{INFOCOM12_HowGoodIsBargained_BlocqOrda}, the authors proposed a new measure called {\em Price of Selfishness} (PoS) to compare the optimal social utility with the social utility obtained at the Nash bargaining solution. However, their definition of PoS is not suitable in our setting to compare $\tilde U$ and $\overline U$. This is because,  in \cite{INFOCOM12_HowGoodIsBargained_BlocqOrda} the problem is defined in a cooperative context in which the regulator determines the actions taken by all the players, i.e, $p_1(\theta)$ and $p_2$, and also the value of $p_d$ that maximizes the product of the utilities of all the players. In our case the problem is not fully cooperative. Bargaining is restricted to the parameter $p_d$ alone. The other parameters are set through competition. Thus in our model bargaining is over a subset of the parameters. We therefore propose an alternative metric called {\em Price of Partial Bargaining} (PoPB), which we define as
\[PoPB=\frac{\tilde{U}}{\overline{U}}.\]
We will next compute the PoPB in the nonneutral regime analyzed in the previous section where $\cp$ shares private information with $\isp_1$ on payment. In the pre bargaining game, side payment $\tilde p_d$ is the maximizer of sum of utilities given by (\ref{eqn:EquilibriumUtiltilityISP1PartialPricing})-(\ref{eqn:EquilibriumUtiltiyPartialPricingCP})
and $p_d^*$ is obtained from (\ref{eqn:SidePayPreBargainOptimization}). The resulting optimal values and the corresponding utilities are cubersome to manupulate. We plot the PoPB in Figure \ref{fig:PoPBPreBargain} as a function of $\tau=\beta/\alpha$ fixing $\alpha=2$. From the figure we note that when $\tau$ is close to $1$, the PoPB is large. When $\tau$ is close to $1$, the demand generated from each $\isp$ is equally sensitive to price set by the competing $\isp$s, in this case pre bargaining leads to poor social utility. However, when $\tau$ is close to $1/2$, PoPB is close to one. This implies that when demand generated from an $\isp$ half as much sensitive as to its own price then the resulting social utility in pre bargaining is close to optimal.When $\tau$ is close to zero, again the pre bargaining results in poor equilibrium social utility.

We note that our definition of PoPB is not appropriate for the post bargaining game. In this game, first optimal side payment is evaluated for a given price of $\isp$s, and then the equilibrium prices are computed. But the similar process of evaluating social utility becomes independent of side payment.


\section{Conclusions}
\label{sec:Conclusion}
In this paper we studied preferential treatment of $\isp$s by $\cp$s through collusions. We modeled a nonneutral behavior in which a $\cp$ shares private about its content through signals. We showed that the $\cp$ may not benefit sharing its private information, whereas $\isp$s always benefit receiving signals.   If the $\cp$ charges the $\isp$s to share its private information, both the $\cp$ and the $\isp$ in collusion may lose, whereas the $\isp$s which do not receive signals may gain.

We also studied two mechanisms based on weighted proportional fairness criteria to set the price (side payments) that $\isp$ pays to the $\cp$
for providing signals. In deciding this side payments we took into account the bargaining power of the players.
We noted that the bargaining power influences players preference for the mechanisms. We also introduced a new performance measure to compare the social utility at equilibrium with the optimal social utility when some parameters are agreed through bargaining and others are set competitively.

\bibliographystyle{IEEEtran}
\bibliography{m2}

\newpage

\section*{appendices}
\subsection{Proof of Theorem \ref{thm:SignalNeutralUtiltiyCompare}}
\begin{IEEEproof}
We begin with the neutral regime with no information. Taking the logarithm of the utility of $\isp_i$ we get
\[\log E[U_{\isp_i}]=\log E[d_i(\theta,p_i,\mathbf{p}_{-i})] + \log p_i\]
Using Assumption \ref{asm:SignalSupermodularity} and monotonicity properties of $d_i(\theta, \mathbf{p})$ given in (\ref{eqn:SignalModel_MonotonicityAssumption}), it is easy to verify that $\log E[d_i(\theta,p_i,\mathbf{p}_{-i})]$ satisfies supermodular property, i.e,
\[\frac{\partial ^2 \log E[d_i(\theta,p_i,\mathbf{p}_{-i})]}{\partial p_i \partial p_j} \geq 0 \;\text{for}\; j\neq i.\]
Then existence of equilibrium follows from Topkis's theorem \cite{Econometrica90_RationalizabilityLearningAnd_MilgromRoberts}.

Using the dominant diagonal property it is easy to verify that for all $i=1,2, \cdots, n$
\[\sum_{j=1}^n \frac{\partial^2 \log E d_i(\theta, \mathbf{p})}{\partial p_i \partial p_j} \leq 0 .\]
Then uniqueness of equilibria follows from \cite{Econometrica90_RationalizabilityLearningAnd_MilgromRoberts}.

Now consider the case with full information. We first note that the demand function for each $\isp$ is separable in $\theta$. Thus, given
$\theta$, $\isp_i$ sets a price that maximizes $U_{\isp}(\theta, p_i(\theta),\mathbf{p}_{-i}(\theta))$ independent of what other $\isp$s set when the signal is different from $\theta$. Hence we can restrict the study of price competition between the $\isp$ for a given $\theta$. It can be easily verified that $d_i(\theta,p(\theta))$ is log-supermodular. Also, by setting $\theta=\gamma$ in Assumption \ref{asm:SignalDominantDiagonal} the condition
\[\sum_{j=1}^n \frac{\partial^2 \log d_(\theta,p(\theta))}{\partial p_i\partial p_j} \leq 0\]
holds for all  $\theta$ and $i$. Then existence and uniqueness follows from Topkis's theorem \cite{Econometrica90_RationalizabilityLearningAnd_MilgromRoberts}.

\end{IEEEproof}

\subsection{Proof of Theorem \ref{thm:ISPUtilityComparison}}
\label{app:ISPUtilityComparison}
\begin{IEEEproof}
For a given $\theta$ the utility of each $\isp$ is quadratic in price. We compute the equilibrium prices by simply solving the best response. A straight forward calculations results in the
following equilibrium utilities when the $\cp$ does not give signal to any of the $\isp$s:
\begin{equation}
\label{eqn:EquilibriumUtilityNoInfo}
E[U_{\isp_1}]=E[U_{\isp_2}]=\frac{\alpha(E[D(\theta)])^2}{(2\alpha-\beta)^2},
\end{equation}
\begin{equation}
\label{eqn:EqulibirumUtilityNoInfoCP}
E[U_{\cp}(\theta,p_1^*,p_2^*)]=\frac{2\alpha E[D(\theta)]}{2\alpha-\beta}p_a.
\end{equation}
Similar calculation results in the following utilities of when the $\cp$ gives signal to both the $\isp$s.
\begin{equation}
\label{eqn:EquilibriumUtilityFullInfo}
E[U_{\isp_1}]=E[U_{\isp_2}]=\frac{\alpha}{(2\alpha-\beta)^2}E[D^2(\theta)],
\end{equation}
\begin{equation}
\label{eqn:EquilibriumUitlityFullInfoCP}
E[U_{\cp}(\theta, p_1^*(\theta),p_2^*(\theta))]=\frac{2\alpha E[D(\theta)]}{2\alpha-\beta}p_a.
\end{equation}
Subtracting the expected utility of the $\isp_1$ in (\ref{eqn:EquilibriumUtilityNoInfo}) from (\ref{eqn:EquilibriumUtilityFullInfo}) we have
\begin{eqnarray*}
\frac{\alpha}{(2\alpha-\beta)^2}(E[D^2(\theta)]-E^2[(\theta)])\\
=\frac{\alpha}{(2\alpha-\beta)^2}Var(D(\theta))\geq 0.
\end{eqnarray*}
Where $Var(D(\theta))$ denotes the variance of the random variable $D(\theta)$.
Now assume that $\cp$ colludes with $\isp_1$ and shares private information only it. Then, the expected utility of $\isp_2$ and $\cp$ at equilibrium can be computed, respectively, as follows:
\begin{equation}
\label{eqn:EquilibriumUtilityPartialInfoISP1}
E[U_{\isp_1}]=\frac{E[D^2(\theta)]}{4\alpha}+ \frac{\beta E^2[D(\theta)](4\alpha-\beta)}{4\alpha (2\alpha-\beta)^2}.
\end{equation}
\begin{equation}
\label{eqn:EquilibriumUtilityPartialInfoISP2}
E[U_{\isp_2}]=\frac{\alpha E^2[D(\theta)]}{(2\alpha-\beta)^2},
\end{equation}
\begin{equation}
\label{eqn:EquilibriumUtilityPartialInfoCP}
E[U_{\cp}]=2\alpha\frac{ E[D(\theta)]}{(2\alpha-\beta)}.
\end{equation}
We now compare the performance of $\isp$ that receives the signaling information with the $\isp$ which do not have this information.
To prove the first claim in the nonneutral regime we compare the expected utility in (\ref{eqn:EquilibriumUtilityPartialInfoISP1}) with the expected utility in (\ref{eqn:EquilibriumUtilityNoInfo}) obtained when both the $\isp$s do not get signaling information. A simple manipulations yields that (\ref{eqn:EquilibriumUtilityPartialInfoISP1}) is larger than (\ref{eqn:EquilibriumUtilityNoInfo}) if and only if $E[D^2(\theta)]\geq E^2[D(\theta)]$, which always holds. \\
To prove the second claim we compare the expected utility in (\ref{eqn:EquilibriumUtilityPartialInfoISP1}) with the expected utility in (\ref{eqn:EquilibriumUtilityFullInfo}) obtained when both the $\isp$s get signaling information. Again, a simple manipulation shows that (\ref{eqn:EquilibriumUtilityFullInfo}) is larger than (\ref{eqn:EquilibriumUtilityPartialInfoISP1}) if and only if $E[D^2(\theta)]\geq E^2[D(\theta)]$ which holds always. \\
The third claim holds by comparing utility of the non-colluding $\isp$ in (\ref{eqn:EquilibriumUtilityFullInfo}) and (\ref{eqn:EquilibriumUtilityPartialInfoISP2}).\\
Finally, the last claim follows by noting that expected utility of the $\cp$ given by in the three cases reo(\ref{eqn:EquilibriumUtilityPartialInfoISP2}) and (\ref{eqn:EquilibriumUtilityNoInfo}) are the same.
\end{IEEEproof}

\subsection{Proof of Proposition \ref{prop:EquilibriumUtiltiyPartialPricing}}
\label{app:EquilibriumUtiltiyPartialPricing}
\begin{IEEEproof}
Best response of $\isp_1$ for a given value of $\theta$ and $p_2$ is
\begin{equation}
\label{eqn:PartialInfoPricingBestResponseISP1}
p_1(\theta)=\frac{D(\theta)+ \beta p_2 + \alpha p_d}{2\alpha}.
\end{equation}
Similarly, the best response of $\isp_2$ for a given strategy profile $\{p_1(\theta): \theta \in \Theta \}$
\[p_2=\frac{E[D(\theta+ \beta p_(\theta))]}{2\alpha}.\]
Solving the above best response equations simultaneously, the equilibrium prices are given by
\begin{eqnarray*}
p_1^*(\theta)&=&\frac{D(\theta)}{2\alpha}+ \frac{\beta E[D(\theta)]}{2\alpha (2\alpha-\beta)} + \frac{2\alpha^2 p_d }{4\alpha^2-\beta^2},\\
p_2&=&\frac{E[D(\theta)]}{2\alpha-\beta} + \frac{\alpha\beta}{4\alpha^2-\beta^2}.
\end{eqnarray*}
Substituting these prices in the utility functions and taking expectation we obtain the equilibrium utility for each player as follows:
\begin{eqnarray*}
\nonumber
\hspace{-.4cm}E [U_{\isp_1}]\hspace{-.4cm} &=& \hspace{-.3cm}\alpha E \left [ \left ( \frac{D(\theta)}{2\alpha}+ \frac{\beta E[D(\theta)]}{2\alpha(2\alpha-\beta)}-\frac{p_d(2\alpha^2-\beta^2)}{4\alpha^2-\beta^2}\right )^2 \right ]\\
\\
E[U_{\isp_2}]\hspace{-.3cm}&=&\hspace{-.3cm}\alpha \left ( \frac{E[D(\theta)]}{(2\alpha-\beta)}+ \frac{\alpha\beta}{4\alpha^2-\beta^2}p_d\right )^2\\
\nonumber
E[U_{\cp}]&=&\frac{2\alpha E[D(\theta)]}{2\alpha-\beta}  -\alpha \frac{(2\alpha^2-\beta^2)}{4\alpha^2-\beta^2}p_d^2 \\
&+&  \left ( \frac{\alpha E[D(\theta)]}{2\alpha-\beta}-\frac{\alpha (2\alpha^2-\beta^2-\alpha \beta)p_a}{4\alpha^2-\beta^2}\right )p_d.
\end{eqnarray*}
The claim follows by comparing the above utilities of the $\isp$s with the utilities given in (\ref{eqn:EquilibriumUtilityPartialInfoISP1})-(\ref{eqn:EquilibriumUtilityPartialInfoISP2}) which corresponds to the case when both the $\isp$s receive signals.
\end{IEEEproof}

\subsection{Proof of Theorem \ref{thm:SignalWithPricingPriceRegion}  }
\label{app:SignalWithPricingPriceRegion}
\begin{IEEEproof}
Collusion with the $\cp$ is beneficial to $\isp_1$ if it can get higher expected utility compared to the case when it does not enter into any agreement. This happens if the expected utility, given in (\ref{eqn:EquilibriumUtiltilityISP1PartialPricing}), is larger than that given in (\ref{eqn:EquilibriumUtilityNoInfo}). Subtracting (\ref{eqn:EquilibriumUtilityNoInfo}) from (\ref{eqn:EquilibriumUtiltilityISP1PartialPricing}) and simplifying, we get the following quadratic equation in $p_d$.
\[f(p_d):=p_d^2-\frac{2E[D](2\alpha+\beta)p_d}{2\alpha^2-\beta^2}+
 \frac{(4\alpha^2-\beta^2)^2Var(D(\theta))}{(2\alpha^2-\beta^2)4\alpha^2}.\]
The roots of this quadratic equation are
\[\frac{(2\alpha+\beta) E[D(\theta)]}{2\alpha^2-\beta^2}\left \{1 \pm \sqrt{1-\frac{(2\alpha-\beta)^2 Var(D(\theta))}{4\alpha^2 E^2[D(\theta)]}}\right \}.\]
Let $x_1$ and $x_2$ denote the smaller and larger root respectively. Note that $f(p_d)$ is a concave function in $p_d$. It takes non negative values outside the interval $[x_1,\;x_2]$. It is easy to verify that for $p_d \geq x_2$ revenue obtained by the colluding $\isp$ is negative. Thus the claim follows by noting that $f(p_d)$ is nonnegative for $p_d\leq x_1$.\\
Similarly, collusion with $\isp_1$ is beneficial to $\cp$ if it can get higher expected utility compared to the case when it does not enter into any agreement. This happens if the expected utility for $\cp$ in collusion, given in (\ref{eqn:EquilibriumUtiltiyPartialPricingCP}), is larger than that given in (\ref{eqn:EqulibirumUtilityNoInfoCP}). Subtracting (\ref{eqn:EqulibirumUtilityNoInfoCP}) from (\ref{eqn:EquilibriumUtiltiyPartialPricingCP}) and simplifying, we get the following quadratic equation in $p_d$.
\[g(p_d):=p_d\left( \frac{ E[D(\theta)]}{2\alpha-\beta}-\frac{(2\alpha^2-\beta^2-\alpha \beta)p_a}{4\alpha^2-\beta^2}-\frac{2\alpha^2-\beta^2}{4\alpha^2-\beta^2}p_d\right ).\]
Now the claim follows by noting that $g(p_d)$ is positive if and only if $p_d$ satisfies the relation (\ref{eqn:CondnPatialPriceCPBenefit}).\\
The last claim can be verified in a similar way by comparing expected utility of $\isp_1$
and $\isp_2$ given in (\ref{eqn:EquilibriumUtiltilityISP1PartialPricing}) and (\ref{eqn:EquilibriumUtiltilityISP2PartialPricing}) respectively.
\end{IEEEproof}

\end{document}